\documentclass[aps,pra,twocolumn,groupedaddress,floatfix,nolongbibliography,nofootinbib]{revtex4-2}
\usepackage[english]{babel}
\usepackage{color}
\usepackage{amsmath}
\usepackage{physics}
\usepackage{soul}
\usepackage{xr-hyper}
\usepackage[hidelinks]{hyperref}
\usepackage{diagbox}
\usepackage{multirow}
\usepackage{amsfonts}
\usepackage{amssymb}
\usepackage{latexsym}
\usepackage{graphicx}
\usepackage{adjustbox}
\usepackage{float}
\usepackage{xr}
\usepackage{caption}
\usepackage{enumitem}
\usepackage{subfigure}
\usepackage{seqsplit}
\usepackage{ytableau}
\usepackage{booktabs}

\usepackage{mathtools}

\newcommand{\al}{\alpha}

\newcommand{\pa}{\partial}

\newcommand{\la}{\lambda}

\newcommand{\rar}{\rightarrow}

\newsavebox\CBox

\begin{document}
	
	
	\title{Estimating the Accuracy of the Variational Energy: The Hydrogen Atom in a Magnetic Field as an Illustration}
	
	\author{J.C.~del~Valle}
	\email{juan.delvallerosales@drake.edu}
	\address{Department of Physics and Astronomy, Drake University, Des Moines, IA 50311, USA}
	
	\begin{abstract}
		For a hydrogen atom  subject to a constant  magnetic field, we report 
		a numerical {realization} of the two-dimensional Non-Linearization Procedure (NLP) to estimate the accuracy of the variational energy associated with a given trial function.
		Relevant equations of the NLP, which resemble  those describing a dielectric medium  with a space-dependent permittivity and charge distribution, are solved numerically with high accuracy  by using an orthogonal collocation method. As an illustration, we consider three different trial functions including the one proposed in 
		Phys. Rev. A \textbf{103}, (2021) 032820.  For a magnetic field strength of $\gamma=1$~a.u., we establish the accuracy of the variational energy  as well as the local accuracy of the trial function. Additionally, the accuracy  of the cusp parameter and the quadrupole moment, found by means of the trial function, is investigated
	\end{abstract}
	
	\maketitle
	\newpage
	
	
	\section*{Introduction}
	The experimental determination of ionization and transition energies of small systems has recently reached unprecedented high accuracy (e.g.,~\cite{Holsch,Clausen,Markus,Lepson}).  Measurements of ionization energies, in particular, provide valuable information on stability, electronic structure, and ground-state energies of ions, atoms, and molecules. When good agreement between experimental and theoretical energies is found, it is possible to refine the value of fundamental constants, as was recently done for the Rydberg constant~\cite{Grinin}.
	
	The calculation of high-precision theoretical energies for small systems is typically accomplished in two steps. The first one is the calculation of a highly-accurate non-relativistic spectrum. The second one usually takes into account finite-mass (if not already considered in the previous step), relativistic, and quantum electrodynamics  corrections via perturbation theory. For example, this two-step procedure was recently used to benchmark the intra\-shell energy transitions with $n=2$  in He~\cite{Yerokhin}.
	
	Most of the high-precision calculations of non-relativistic energies are rooted in the Variational Method (VM). This method provides an upper bound (also known as variational energy) for the ground state energy, and  also for excited states  when the \hbox{Hylleraas}–Undheim–MacDonald theorem~\cite{HUM} applies. The underlying ingredient behind the VM is the trial function: a parameter-dependent approximation of the unknown wave function. The presence of variational parameters allows searching for the optimal configuration to find the lowest upper bound to the ground state that the trial function can provide.
	
	Unfortunately, the VM does not provide information about the closeness between the upper bound and the exact ground-state energy.
	To overcome this drawback, the formalism of lower bounds was developed in~\cite{Temple}. In this way, both the upper and lower bounds define the domain where the exact theoretical energy is located. Lower bounds turned out to be considerably less accurate than upper bounds, with a rate of convergence several orders of magnitude slower than the rate obtained for the upper bound~\cite{Martinazzo}. Nevertheless, recent refinements of lower  bounds have reached  fast convergence and high accuracy for some atoms such as He and Li;  see~\cite{Nakashima} and~\cite{Ireland}. By construction, the calculation of lower bounds is focused on trial functions that contain linear parameters only.  Furthermore, this consideration does not deliver information about the accuracy of the trial function.
	
	As an alternative to the calculation of lower bounds, we can use the connection between the VM and Perturbation Theory (PT) to estimate the accuracy of the variational energy and, most importantly, the trial function. This connection establishes that calculating the variational energy is equivalent to the computation of the first two lowest corrections in PT; see, e.g.,~\cite{Epstein,Turbiner1984}. In this framework, the trial function represents the zeroth-order correction in PT of the exact wave function. Thus, as long as PT is convergent~\cite{Turbiner1984}, the accuracy of the variational energy and the trial function can be estimated through the calculation of higher-order corrections. Consequently, the accuracy of any quantity derived from the trial function, such as expectation values, may be estimated as well.
	
	An accurate computation of the corrections is a computationally expensive task using the standard formulation of PT, also known as Rayleigh-Schr\"odinger PT~\cite{Schiff}. The main drawback of that formulation is the need of the unperturbed spectrum (at least in approximate form). Alternatively, we can use the Non-Linearization Procedure (NLP)~\cite{Turbiner1980}, which is an efficient method  that only requires the unperturbed wave function of the state of interest as input. The NLP has been extensively used to estimate the accuracy of variational energies and trial functions for various systems described by a single degree of freedom, e.g., the one-dimensional quantum anharmonic oscillator~\cite{book}. To the best of our knowledge, calculations performed for multidimensional systems described by more than one degree of freedom, have not yet been reported.
	
	The present work represents an attempt to estimate the accuracy of the variational energy and trial function via the NLP in a system described by more than one degree of freedom. We employ a collocation method to solve all relevant partial differential equations of the NLP. As an illustration, and a step toward considering more complicated systems, we applied the formalism to a hydrogen atom in its ground state subjected to a constant magnetic field, which is one of the simplest atomic systems described by two degrees of freedom. We choose three relevant trial functions found in the literature proposed by Yafet \textit{et al}.~\cite{Yafet}, Turbiner~\cite{Turbiner-Zeeman}, and  del Valle-Turbiner-Escobar (dVTE)~\cite{TB,TB-erratum}. After Yafet \textit{et al.}\ reported their trial function, Turbiner constructed a locally accurate trial function. We also consider the compact trial function recently proposed in~\cite{TB}, whose construction was based on asymptotic analysis, the semi-classical expansion, and PT in powers of the strength of the magnetic field $\gamma$. Its high accuracy enables the investigation of finite-mass effects on the energy in the range $\gamma\in[0,10\,000]$~a.u.. While numerical results suggested the local accuracy of the trial function~\cite{TB}; we justify this property in the present follow-up work.
	
	This manuscript is organized as follows. We start by describing the connection between VM and PT in section~\ref{VM}. Then, in section \ref{Generalities}, we recapitulate the essentials of the  NLP. In section \ref{Application}, we describe a suitable numerical realization of the NLP for the system of interest. Numerical results are reported in sections \ref{Results} and \ref{section:expec}. We finish with conclusions in \ref{Conclusions}. Atomic units are used throughout this text.
	
	
	\section{Variational Method and Perturbation Theory}
	\label{VM}
	The VM lies in the variational principle, which establishes
	\begin{equation}
		E \leq \dfrac{\int_\Omega\psi_0^*\hat{H}\psi_0\,dV}{\int_\Omega|\psi_0|^2\,dV} := E_{var}[\psi_0],
		\label{VP}
	\end{equation}
	where $E$ is the exact ground state energy and $\psi_0$ is a square-integrable (trial) function. The Hamiltonian operator of the system, $\hat{H}$, is assumed to be a bounded-from-below operator.
	The integration is carried out over the domain $\Omega$ on which the corresponding Schr\"odinger equation is defined. The upper bound is called the variational energy; it is denoted by $E_{var}[\psi_0]$ in (\ref{VP}). The equality between $E$ and $E_{var}[\psi_0]$ holds if $\psi_0$ is the exact ground state eigenfunction.
	
	Let $\hat{H}$  be the Hamiltonian given by
	\begin{equation}
		\hat{H} = -\frac{1}{2}\Delta + V,
		\label{H}
	\end{equation}
	where $\Delta$ denotes the Laplacian and $V$ is the potential energy. While the Schr\"odinger equation
	\begin{equation}
		\hat{H}\psi =  E\psi; \qquad \int_\Omega|\psi|^2\,dV < \infty
		\label{interest}
	\end{equation}
	can be rarely solved exactly for a given $V$, the \textit{inverse} problem is always solvable: for a given square-integrable nodeless (trial) function $\psi_0$, one can find the potential $V_0$ for which $\psi_0$ is the ground-state wave function of the corresponding Schr\"odinger equation. This potential is given by
	\begin{equation}
		V_0 - E_0 = \frac{1}{2}\frac{\Delta\psi_0}{\psi_0} .
		\label{unperturbed}
	\end{equation}
	Then, from  (\ref{VP}), (\ref{H}), and (\ref{unperturbed}), we have
	\begin{equation*}
		E_{var}[\psi_0] = E_0 + \la E_1;\qquad \la=1 , 
	\end{equation*}
	where
	\begin{equation*}
		E_1 = \dfrac{\int_\Omega\psi_0^*(V-V_0)\psi_0\,dV}{\int_\Omega|\psi_0|^2\,dV}.
	\end{equation*}
	This equation reveals that the calculation of $E_{var}[\psi_0]$ is equivalent to computing the two-lowest order corrections in PT, where $V-V_0$ represents the perturbation potential to the unperturbed~$V_0$~\cite{Epstein}. Thus, $E_0$ and  $\psi_0$ play the role of the zeroth-order corrections for the energy and wave function, respectively.
	
	With higher-order corrections $E_n$, we can construct
	\begin{equation}
		E = \sum_{n=0}^\infty\lambda^ n\,E_n.
		\label{PT-Energy}
	\end{equation}
	If the radius of convergence of the series is sufficiently large to have $\lambda=1$ inside, the partial sums of (\ref{PT-Energy}) may lead to highly accurate estimates of the ground-state energy~\cite{book}. In particular, when the second-order correction $E_2$ is the leading one, it measures the accuracy of the variational energy~\cite{Turbiner1984}. Similarly, the closeness of $\psi_0$ to the exact solution $\psi$ can be established by calculating the corrections to the unperturbed wave function $\psi_0$. Remarkably, a finite radius of convergence of the series (\ref{PT-Energy}) is obtained if $\psi_0$ is such that
	\begin{equation}
		\label{guarantee}
		\left|\frac{V-V_0}{V_0}\right| \lesssim \text{const}
	\end{equation}	
	for $|\boldsymbol{x}|>R$ for sufficiently large~$R$. When (\ref{guarantee}) is fulfilled, the potential $V-V_0$ is a subordinate perturbation of $V_0$~\cite{Turbiner1980,Turbiner_Beyond}.
	
	
	\section{Overview of the Non-Linearization Procedure}
	\label{Generalities}
	Consider the Schr\"odinger equation (\ref{interest}) defined over the domain $\Omega$ with the Hamiltonian (\ref{H}). We assume that $\psi$ denotes the exact and unknown ground-state wave function. The function $\psi$ is characterized by being nodeless, i.e., it does not vanish inside $\Omega$. This property suggests taking the exponential representation of the function, namely
	\begin{equation}
		\psi = \exp({-\Phi}) .
		\label{ExpRep}
	\end{equation}
	The \textit{phase} of the wave function, $\Phi$, is real and non-singular. One can verify that $\Phi$ obeys the non-linear (partial) differential equation
	\begin{equation}
		\Delta\Phi - \nabla\Phi\cdot\nabla\Phi = 2(E - V) ,
		\label{Riccati}
	\end{equation}
	where $\nabla$ denotes the gradient. We consider a potential $V$  split into two components: ${V = V_0 + \lambda V_1}$, where $\lambda$ is a parameter.
	
	Together with (\ref{PT-Energy}), the solution of (\ref{Riccati}) is taken in the form of a power series in $\lambda$, namely
	\begin{equation}
		\Phi = \sum_{n=0}^\infty\la^n\Phi_n.
		\label{PTseries}
	\end{equation}
	The $n$th-order corrections, $E_n$ and $\Phi_n$, are defined by the linear (partial) differential equation
	\begin{equation}
		\nabla\cdot(\psi_0^2\,\nabla\Phi_n)= 2(E_n - Q_n)\psi_0^2,\qquad \psi_0 = \exp(-\Phi_0) ,
		\label{electric}
	\end{equation}
	where
	\begin{equation*}
		Q_1 = V_1,\qquad Q_{n} = -\frac{1}{2}\sum_{k=1}^{n-1}\nabla \Phi_{n-k}\cdot\nabla \Phi_{k} ,\qquad n>1 .
	\end{equation*}
	For bound states, the requirement of vanishing probability current at the boundary of $\Omega$~\cite{Turbiner1984}
	\begin{equation}
		\int_{\partial\Omega}\psi_0^2\,\nabla\Phi_n\cdot d\vec{S}=0,
		\label{condition}
	\end{equation}
	must be imposed on (\ref{electric}). Consequently, from an integration of (\ref{electric}) over $\Omega$, we obtain  the $n$th-order energy correction in integral form,
	\begin{equation}
		E_n = \dfrac{\int_{\Omega}Q_{n}\,\psi_0^2\,dV}{\int_{\Omega}\psi_0^2\,dV}.
		\label{En}
	\end{equation}
	
	In its one-dimensional version, (\ref{electric}) can always be solved, leading to $\Phi_n$ in integral form. In this case, the numerical realization of the NLP is straightforward and can be used to calculate high orders in PT~\cite{book}. In contrast, numerical methods  are required in the multidimensional case since the exact solution of (\ref{electric}) is not available in general. In the present work, we use a collocation method to solve (\ref{electric}). The choice of this method is motivated by the connection between (\ref{electric}) and the equation that describes  the electrostatics of a dielectric medium  with a space-dependent permittivity and charge distribution. To exhibit this connection, we define three auxiliary objects:
	\begin{equation*}
		\varepsilon = \psi_0^2 ,\qquad \boldsymbol{E}  = \nabla\Phi_n ,\qquad\rho = (E_n-Q_n)\psi_0^2 ,
		\label{poisson}
	\end{equation*}
	called permittivity, electric field, and density of free charge, respectively. In terms of these, equation (\ref{electric})  can be expressed as
	\begin{equation}
		\nabla\cdot(\varepsilon\,\boldsymbol{E})=\rho,
		\label{gauss}
	\end{equation}
	which is the familiar Gauss's law  describing a dielectric medium characterized by a space-dependent permittivity.
	Thus, calculating the corrections $\Phi_n$ and $E_n$ is equivalent to solving an electrostatic problem for each $n$. For equations of the type (\ref{gauss}), collocation methods have shown to be adequate tools to solve them with high accuracy; see, e.g.,~\cite{Dang-Vu,Chen}.
	
	
	\section{Zeeman Effect in the Hydrogen atom}
	\label{Application}
	We now apply the NLP to a physically relevant atomic system described by two degrees of freedom. We consider the hydrogen atom in its ground state subjected to a constant magnetic field, resulting in the celebrated (quadratic) Zeeman effect.
	
	\subsection{Main Equations}
	The Hamiltonian operator is conveniently written in three-dimensional spherical coordinates:
	\begin{equation}
		\hat{H} =  -\frac{1}{2}  \Delta  + V ,\qquad V = -\ \frac{1}{r} + \frac{\gamma^2}{8}r^2\sin^2 \theta ,
		\label{HinB}
	\end{equation}
	where  $\Delta$ is three-dimensional Laplace operator, while $\gamma$ denotes the strength of the magnetic field. The following assumptions are behind~(\ref{HinB}): i)~the infinitely-massive proton is located at the origin of the coordinate system; ii)~the  magnetic field $\boldsymbol{B}=\gamma\,\hat{\boldsymbol{z}}$ is described by the symmetric gauge; iii) spin and other relativistic effects are neglected.
	
	Since the ground-state wave function  exhibits a {$(r,\theta)$-dependence} only, (\ref{electric}) takes the form
	\begin{align}
		&\frac{1}{r^2}\pa_r(r^2\pa_r\Phi_n) + \frac{1}{r^2}\pa_{u}\left((1-u^2)\pa_u\Phi_n\right)
		\nonumber \\
		&-2\left[\pa_r\Phi_0\,\pa_r\Phi_n + \frac{1-u^2}{r^2}\pa_u\Phi_0\,\pa_u\Phi_n\right] = E_n - Q_n,
		\label{eqPhi}
	\end{align}
	with
	\begin{align*}
		Q_1 =& -\frac{1}{r}+\frac{\gamma^2}{8}r^2(1-u^2) - V_0,\\
		Q_n =& -\frac{1}{2}\sum_{k=1}^{n-1}\left[\pa_r\Phi_{n-k}\,\pa_r\Phi_{k} + \frac{1-u^2}{r^2}\pa_u\Phi_{n-k}\,\pa_u\Phi_{k}\right],
	\end{align*}	
	 see \cite{TB}, where $u=\cos\theta$.
	In the set of coordinates $\{r,u\}$, the boundary condition (\ref{condition}) reads
	\begin{equation}
		\lim_{R \to \infty} \int_{-1}^{1} e^{-2\Phi_0(R,u)}\pa_r\Phi_n(r,u)\bigg\rvert_{r = R}\!\!\!\!\!\!\!\!du = 0.
		\label{Boundary}
	\end{equation}	
	Since the exact wave function is even in the sense ${\psi(r,u) = \psi(r,-u)}$, so are the corrections $\Phi_n(r,u)$. Consequently, we can reduce the $u$-domain from $u\in[-1,1]$ to $u\in[0,1]$.
	
	
	\subsection{Collocation Method}
	We assume that any correction $\Phi_n(r,u)$ has the representation
	\begin{equation}
		\Phi_n(r,u) = \sum_{k=1}^{N_r}\sum_{l=1}^{N_u/2}c^{(n)}_{kl}f_k(r)g_l(u),
		\label{expansion}
	\end{equation}
	where  $c_{kl}^{(n)}$ are coefficients, and the functions $f_k(r)$ and $g_l(u)$ are polynomials of degree $N_r$ and (even) $N_u$, respectively. Without loss of generality, we set $\Phi_n(0,u)=0$ as the normalization of any correction. Therefore, we demand $f_k(0)=0$ in (\ref{expansion}). The explicit form of the polynomials is presented below. The exponential decay of $\psi_0=e^{-2\Phi_0(r,u)}$ and the polynomial growth of $|\Phi_n(r,u)|$ ensures that the boundary condition (\ref{Boundary}) is fulfilled.
	
	Consider $\{r_i\}_{i=1}^{N_r}$ and $\{u_l\}_{l=1}^{N_u/2}$ as collocation points. The polynomials  $f_k(r)$ and $g_l(u)$ are constructed in such a way that they obey
	\begin{equation}
		f_k(r_i) = \delta_{ik} ,\qquad g_l(u_j) = \delta_{lj} ,
		\label{Kronecker}
	\end{equation}
	where  $\delta_{ij}$ denotes the Kronecker delta. After substituting the expansion (\ref{expansion}) into (\ref{eqPhi}), and demanding that the equation is satisfied at each point $(r_i,u_j)$, it is found that a system of linear equations determines the coefficients~$c_{kl}^{(n)}$.  Specifically,
	\begin{gather}
		\sum_{k=1}^{N_r}\sum_{l=1}^{N_u/2} \left\{A_{ik}^{(j)}\,\delta_{jl} + B_{jl}^{(i)}\,\delta_{ik}\right\}\,
		c_{kl}^{(n)} = E_n - Q_{n}(r_i,u_j) ,
		\label{numerical}
	\end{gather}
	where
	\begin{equation}
		A_{ik}^{(j)}=
		f_k''(r_i) + 2f_k'(r_i)\left[\frac{1}{r_i} -\pa_r\Phi_0(r_i,u_j)\right]
		\label{radial}
	\end{equation}
	and
	\begin{align}
		B_{jl}^{(i)} =\, &(1-u_j^2)g_k''(u_j)
		\nonumber \\
		&-2\left[u_l-(1-u_j^2)\pa_u\Phi_0(r_i,u_j)\right]g_l'(u_j).
		\label{angular}	
	\end{align}
	
	We now provide the explicit form of $f_k(r)$ and $g_l(r)$. The function $f_k(r)$ is constructed in terms of the Laguerre polynomial of $L_{N_r}(r)$ whose $N_r$ zeros are the collocation points. Concretely,
	\begin{equation}
		f_k(r) = \frac{r}{r_k}\frac{ \,L_{N_r}(r)}{(r-r_k)L_N'(r_k)}\ ,\qquad L_{N_r}(r_k) = 0\ ,
		\label{f(r)}
	\end{equation}
	where $k=1,\ldots,N_r$. On the other hand, we use the Legendre polynomials $P_{N_u}(u)$ and its zeros $u_l$ as collocation points to define
	\begin{equation}
		g_l(u) = h_l(u) + h_{-l}(u) , \qquad h_l(u) = \frac{ P_{N_u}(u_l)}{(u-u_l)P_N'(u_l)} ,
		\label{g(u)}
	\end{equation}
	with
	\begin{equation*}
		P_{N_u}(u_l) = 0 ,\qquad 0<  u_l< 1 ,\qquad u_l = -u_l ,
	\end{equation*}
	where $l=1,\ldots,N_u/2$. The functions (\ref{f(r)}) and (\ref{g(u)}) satisfy the condition~(\ref{Kronecker}). Furthermore, $g_l(u)$ is an even function.
	Using the polynomials $L_{N_r}(r)$ and $P_{N_u}(u)$ leads to compact expressions for the derivatives required by  (\ref{radial}) and (\ref{angular}).
	Explicit expressions are found in Appendix~\ref{Aformulas}.
	
	After solving (\ref{numerical}), the integrals in (\ref{En}) are calculated via the Gauss quadrature associated with the collocation points. Finally, a parameter $R$ is used to scale the collocation points ($r_i\rar R\,r_i$) such that the boundary condition (\ref{Boundary}) is satisfied with high accuracy.
	
	
	\section{Results}
	\label{Results}
	We consider a magnetic field strength $\gamma=1$ and study the accuracy provided by three different functions written in the form $\psi_0=e^{-\Phi_0}$.
	Since all of them satisfy~(\ref{guarantee}), the convergence of~(\ref{PT-Energy}) is guaranteed.
	
	
	\subsection{Yafet \textit{et al.}  function \cite{Yafet}} 
	According to~\cite{Garstang}, Yafet \textit{et al.}~\cite{Yafet} reported an early trial function to study the Zeeman effect on the hydrogen atom. The phase of this trial function has the form
	\begin{equation}
		\label{Yafet}
		\Phi_0 = a r^2(1+b\,u^2) ,
	\end{equation}
	where $a=0.55050$ and $b=-0.25137$ are the optimal parameters at $\gamma=1$. The variational energy results in ${E_{var} = -0.256\,018}$.
	
	Table~\ref{table:Yafet} shows the first two corrections to the variational energy for different values of $N_r$, $N_u/2$, and~$R$. The convergence in $E_2$ and $E_3$ to five significant digits is reached with $15\times15$ collocation points \hbox{($N_r=N_u/2=15$}). However, $2\times2$ collocation points provide the correct order of magnitude of the first two corrections. From the numerical results, we observe that $E_2$ provides the  dominant contribution to the deviation between the variational and the exact energy.
	
	As was verified several times for one-dimensional anharmonic oscillators~\cite{book},  the ratio  $|E_2/E_3|$ provides an accurate estimate of the rate of convergence of the series for $E$~(\ref{PT-Energy}). A ratio $\left|E_{3}/E_2\right|\sim 2.5\times10^{-1}$ suggests that the series~(\ref{PT-Energy}) is slowly convergent.  This can be seen from the partial sum of the first four energy corrections also shown in Table~\ref{table:Yafet}.
	
	\begin{table*}[t!]
		\caption{Corrections and their partial sum in NLP for the hydrogen atom in a magnetic field using (\ref{Yafet}) as phase $\Phi_0$ for $\gamma=1$. The result of the exact energy, $E_{exact}$, was taken from~\cite{TB}. All numbers are rounded to the digits given.}
	{\setlength{\tabcolsep}{0.4cm}
		\begin{tabular}{|ccccc|c|}
			\hline
			\hline
			\rule{0pt}{4ex}		
			$N_r$ & $N_u/2$&$R$ & $E_2$&$E_3$&$E_0+E_1+E_2+E_3$ \\[5pt]
			\hline
			\rule{0pt}{4ex}
			2     & 2       &1    &$-4.3967\times10^{-2}$&$-1.4337\times10^{-2}$ &$-0.314\,322$ \\[5pt]
			5     & 5       & 1  &$-3.8651\times10^{-2}$ &$-1.0069\times10^{-2}$   &$-0.304\,737$ \\[5pt]
			10    & 10      &1/2   &$-5.3132\times10^{-2}$ & $-1.2749\times10^{-2}$  &$-0.322\,527$\\[5pt]
			15    & 15      &1/3   &$-5.3286\times10^{-2}$ &$-1.3109\times10^{-2}$ &$-0.322\, 413$ \\[5pt]
			20    & 20      &1/4&$-5.3286\times10^{-2}$&$-1.3109\times10^{-2}$  &	$-0.322\,413$			\\[5pt]
			\hline
			\hline
			\multicolumn{4}{c|}{}&\rule{0pt}{4ex} $E_{exact}$&$-0.331\,169$\\[5pt]
			\cline{5-6}
	\end{tabular}}
	\label{table:Yafet}
\end{table*}

\begin{table*}[t!]
	\caption{Corrections and their partial sum in NLP for the hydrogen atom in a magnetic field using (\ref{Turbiner}) as phase $\Phi_0$ for $\gamma=1$. The result of the exact energy, $E_{exact}$, was taken from~\cite{TB}. All numbers are rounded to the digits given.}
{\setlength{\tabcolsep}{0.4cm}
	\begin{tabular}{|ccccc|c|}
		\hline
		\hline
		\rule{0pt}{4ex}		
		$N_r$ & $N_u/2$& $R$&$E_2$&$E_3$&$E_0+E_1+E_2+E_3$ \\[5pt]
		\hline
		\rule{0pt}{4ex}
		2     & 2      &1&$-1.004\times10^{-2}$&$9.91\times10^{-4}$&$-0.338\,610$\\[5pt]
		5     & 5      &1/2&$-1.733\times10^{-3}$ &$2.64\times10^{-4}$&$-0.331\,027$\\[5pt]
		10    & 10      &1/5&$-1.794\times10^{-3}$&$2.51\times10^{-4}$&$-0.331\,102$\\[5pt]
		15    & 15      &1/8 & $-1.794\times10^{-3}$&$2.51\times10^{-4}$&$-0.331\,102$\\[5pt]
		20    & 20      &1/10&$-1.794\times10^{-3}$&$2.51\times10^{-4}$&$-0.331\,102$				\\[5pt]
		\hline
		\hline
		\multicolumn{4}{c|}{}&\rule{0pt}{4ex} $E_{exact}$&$-0.331\,169$\\[5pt]
		\cline{5-6}
\end{tabular}}
\label{table:Turbiner}
\end{table*}


\subsection{Turbiner's function \cite{Turbiner-Zeeman} }

The simplest function that matches the asymptotic behavior of the phase between the weak and ultra-strong regimes was proposed in~\cite{Turbiner-Zeeman}. The phase of this trial function reads
\begin{equation}
\label{Turbiner}
\Phi_0 = a \,r + b \,r^2(1-u^2) ,
\end{equation}
where $a = 1.032370$ and $b = 0.115932$ are optimal for $\gamma=1$ and lead to
${E_{var}  = -0.329\,558.}$

The value of the corrections $E_2$ and $E_3$ is presented in Table \ref{table:Turbiner} for different values of $N_r$, $N_u/2$, and~$R$. Convergence in the first six decimal  digits  of $E_2$ and $E_3$ is achieved with $10\times10$ collocation points. We observe that $E_2$ provides the dominant contribution to the deviation between the variational and the exact energy. The ratio $\left|E_3/E_2\right| \sim 1.4\times10^{-1}$ indicates that the convergence is still slow, but twice as fast compared to the one provided by~(\ref{Yafet}). This can be seen in the partial sum of the first four energy corrections shown in Table~\ref{table:Turbiner}.


\subsection{dVTE function \cite{TB}}

Based on asymptotic analysis, a semi-classical consideration, and perturbation theory, an approximation for the phase was recently proposed in~\cite{TB}. Defining ${\rho^2= r^2\,(1-u^2)}$, the trial function is written as
\begin{align}
\label{approximatephase}
\Phi_0 =
&
-\alpha_0 + \frac{\al_0\,+\,\al_1\,r\,+\,\al_2\,r^2\,+\,\al_3\,\rho^2+\al_4\,\rho^2\,r}
{\sqrt{1  + \beta_1\,r + \beta_2\,r^2+\beta_3\,\rho^2}}\nonumber \\
&
+ \log(1 + \beta_1\,r + \beta_2\,r^2 + \beta_3\,\rho^2),
\end{align}
where $\{\alpha_0,\alpha_1,\alpha_2,\alpha_3,\alpha_4,\beta_1,\beta_2,\beta_3\}$ are variational parameters. Their optimal value  was already established for magnetic field strengths $\gamma\in[0,10\,000]$~\cite{TB}. In this domain, the corresponding trial function consistently provides a relative error in the energy of order $10^{-6}$ or less. For $\gamma=1$, the optimal parameters are presented in Table~\ref{DelValleOptimal}.  They result in 	${E_{var}=-0.331\,168\,829\,376}$.

\begin{table*}[t!]	
\caption{Optimal variational parameters for the phase (\ref{approximatephase}) for $\gamma=1$.}	
{\setlength{\tabcolsep}{0.3cm}
	\begin{tabular}{|c|cccccccc|}
		\hline
		\hline
		\rule{0pt}{4ex}	
		$\gamma$  & $\alpha_0$ & $\alpha_1$ & $\alpha_2$ & $\alpha_3$ & $\alpha_4$ & $\beta_1$ & $\beta_2$ & $\beta_3$ \\[5pt]
		\hline
		\rule{0pt}{4ex}			
		1               &3.23088            &1.13922           &0.12544            &0.06107            & 0.04423          &0.22745           &0.00960           & 0.02480          \\[5pt]
		\hline
		\hline
	\end{tabular}
}
\label{DelValleOptimal}
\end{table*}

\begin{table*}[t!]
\caption{Corrections and their partial sum in NLP for the hydrogen atom in a magnetic field using (\ref{approximatephase}) as phase $\Phi_0$ for $\gamma=1$. The result of the exact energy, $E_{exact}$, was taken from~\cite{TB}. All numbers are rounded to the digits given.}
{\setlength{\tabcolsep}{0.4cm}
\begin{tabular}{|ccccc|c|}
	\hline
	\hline
	\rule{0pt}{4ex}		
	$N_r$ & $N_u/2$&$R$ & $E_2$&$E_3$&$E_0+E_1+E_2+E_3$ \\[5pt]
	\hline
	\rule{-2pt}{4ex}
	2     & 2       &  1  &$-3.7430\times10^{-7}$ &$\ \ \,  3.980\times10^{-11} $&$-0.331\,169\,203\,635$\\[5pt]
	5     & 5       &  1 &$-9.4845\times10^{-8}$ & $-1.790\times10^{-10}$   &$-0.331\,168\,924\,400 $\\[5pt]
	10    & 10      & 1/2  &$-6.6613\times10^{-8}$ &$-3.074\times 10^{-10}$   &$-0.331\,168\,896\,297$\\[5pt]
	12& 10&1/3&$-6.7045\times10^{-8}$&$-3.077\times10^{-10}$&$-0.331\,168\,896\,730$\\[5pt]
	15    & 15      & 1/3  &$-6.7042\times10^{-8}$ &$-3.077\times10^{-10}$ &$-0.331\,168\,896\,726$ \\[5pt]
	20    & 10      &1/4&$-6.7042\times10^{-8}$&$-3.078\times10^{-10}$& $-0.331\,168\,896\,726$				\\[5pt]
	30    & 15  &1/2   &$-6.7042\times10^{-8}$  &$-3.078\times10^{-10}$ &$-0.331\,168\,896\,726$ \\[5pt]
	30    & 30  &1/2   &$-6.7042\times10^{-8}$ &$-3.086\times10^{-10}$ &$-0.331\,168\,896\,727$ \\[5pt]
	\hline
	\hline
	\multicolumn{4}{c|}{}&\rule{0pt}{4ex} $E_{exact}$&$-0.331\,168\,896\,733$\\[5pt]
	\cline{5-6}
	\end{tabular}}
	\label{table:DelValle}
\end{table*}

The value of the corrections $E_2$ and $E_3$ is presented in Table \ref{table:DelValle} for different values of $N_r$, $N_u/2$, and~$R$. Note that $10\times10$ collocation points are sufficient to obtain  corrections with two reliable significant digits. The largest number of collocation points we used is $30\times30$. In this case, the value of $E_3$ deviates from the already converged results using fewer points (e.g., $30\times15$).  This suggests that  $30\times30$ collocation points are optimal for $\Phi_1(r,u)$, but not for $\Phi_3(r,u)$.
The ratio $\left|E_{3}/E_2\right|\sim5.1\times10^{-3}$ indicates a rate of convergence that is at least 100 times faster compared to the  rates provided  by the functions~(\ref{Yafet}) and~(\ref{Turbiner}).

Based on the results for $E_2$ shown in Table \ref{table:DelValle}, we can now include an uncertainty in the value of the ground-state energy. This results in
\begin{equation*}
E = -0.331\,168\,83(7)
\end{equation*}	
and also in the refined value
\begin{equation*}
E = -0.331\,168\,896\,4(3)
\end{equation*}	
if one considers $E_3$.

\begin{figure*}[t!]
\includegraphics[width=\textwidth]{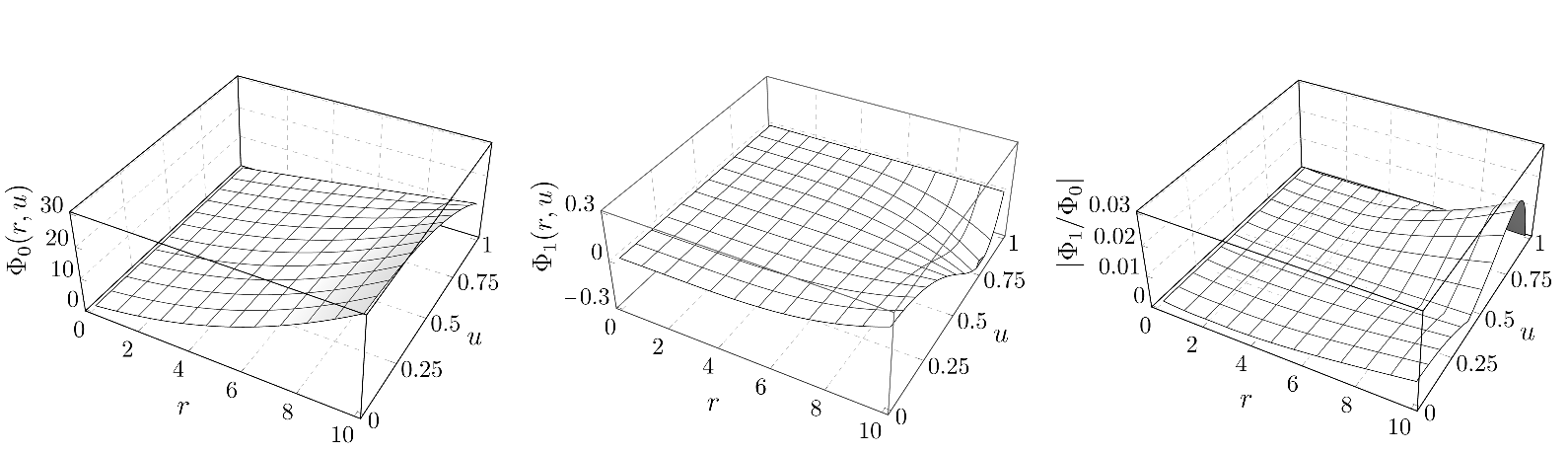}
\caption {Plot of the first two corrections $\Phi_0(r,u)$, $\Phi_1(r,u)$, and the absolute value of their ratio, respectively.}
\label{Fig:corrections}
\end{figure*}

\begin{figure*}[t!]
\includegraphics[scale=0.7]{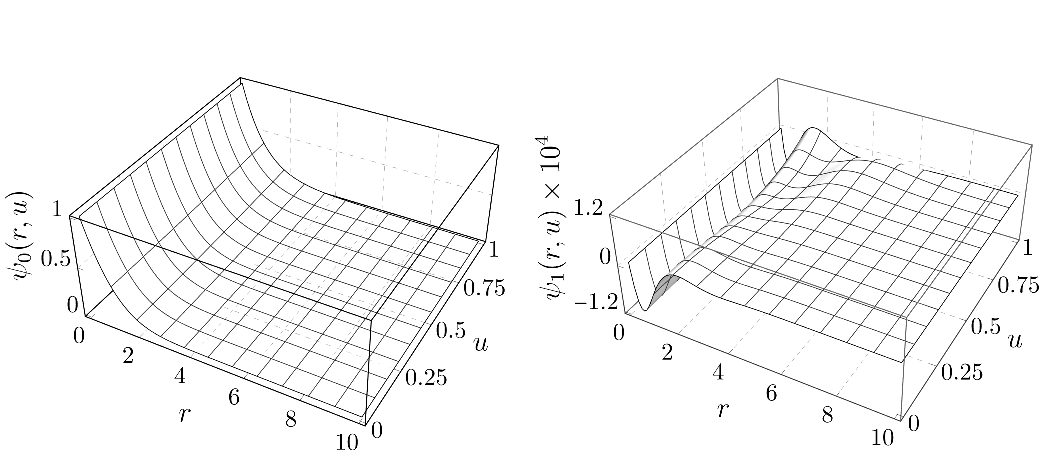}
\caption{Plot of the first two corrections in expansion (\ref{exact}) with phase (\ref{approximatephase}) as the zeroth-order correction. Note the factor ($10^{4}$) used to multiply $\Psi_1(r,u)$ and to make the scales comparable.}
\label{Fig:corrections-psi}
\end{figure*}

Figure~\ref{Fig:corrections} shows plots of the zeroth- and first-order corrections for the phase on ${r\in[0,10]}$. We observe that the correction $\Phi_1(r,u)$ is small compared to $\Phi_0(r,u)$, as can be confirmed by noting the ratio ${|\Phi_1/\Phi_0|\sim0.03}$ (cf.~Fig.~\ref{Fig:corrections}.
Using the corrections $\Phi_0(r,u)$ and $\Phi_1(r,u)$, we can investigate the difference between the trial function and the exact one. From (\ref{ExpRep}) and (\ref{PTseries}), we construct the expansion of the exact wave function in powers of $\la$, resulting in
\begin{equation}
\psi = \psi_0 + \la\psi_1 + \mathcal{O}(\la^2) ,
\label{exact}
\end{equation}
where $ \psi_0 = e^{-\Phi_0}$ and $\psi_1 = -\Phi_1e^{-\Phi_0}$. Plots of $\psi_0(r,u)$ and $\psi_1(r,u)$ are presented in Fig.~\ref{Fig:corrections-psi}.
According to these plots,
\begin{equation}
\left|\psi_{exact}(r,u)-\psi_0(r,u)\right|\approx|\psi_1(r,u)|\lesssim1.2\times10^{-4}
\label{accuracy}
\end{equation}	
for all values of $r$ and $u$. This fact is corroborated by the second-order correction $\psi_2$ (not shown, but $|\psi_2|\lesssim2\times10^{-6}$) ensures that
the inequality~(\ref{accuracy}) holds).
Consequently, we have verified that the trial function proposed in~\cite{TB} is a locally accurate approximation of the exact ground-state wave function. This feature guarantees, in particular, that all expectation values calculated via the trial function are highly accurate.


\section{Accuracy of Expectation Values}
\label{section:expec}

Let us consider the cusp parameter and the magnetic quadruple moment,
\begin{equation}
\label{expectation}
\mathcal{C} =  \frac{{\langle\psi|\delta(\vec{r})\frac{\pa}{\pa r}|\psi\rangle}}{{\langle\psi|\delta(\vec{r})|\psi\rangle}}
,\qquad
Q=3\langle\psi| r^2u^2|\psi\rangle-\langle\psi | r^2|\psi\rangle ,
\end{equation}
respectively. The presence of the magnetic field does not break the cusp condition~\cite{TB}. Hence,  $\mathcal{C}\equiv-1$ for the exact ground-state wave function. In turn, the magnetic field induces a non\-zero quadrupole moment on the hydrogen atom~\cite{Turbiner-Zeeman}.

By means of the expansion (\ref{exact}), we construct the series
\begin{equation*}
\mathcal{C}=\sum_{n}^\infty\mathcal{C}_{n=0}\la^n ,\qquad  Q=\sum_{n=0}^\infty Q_{n}\la^n  .
\end{equation*}
The replacement of $\psi$ by $\psi_0$ in (\ref{expectation}) defines the zeroth-order correction of $\mathcal{C}$ and $Q$ denoted by $\mathcal{C}_0$ and $Q_0$, respectively. If the series are convergent, higher-order corrections are used to estimate the accuracy of the zeroth-order correction $\mathcal{C}_0(Q_0)$, with special interest for the first-order correction $\mathcal{C}_1(Q_1)$.

Table \ref{table:expectation} shows $\mathcal{C}_{i=0,1}$, and $Q_{i=0,1}$ for  the three  functions considered in the previous Section; see (\ref{Yafet}), (\ref{Turbiner}), and (\ref{approximatephase}). Concerning the cusp parameter, Yafet's trial function gives $\mathcal{C}_0 = 0$. However, the first-order correction $\psi_1$ rectifies the wrong  behavior of $\pa_r\psi_0$ at $r=0$. Thus, the corrected value $\mathcal{C}_0+\mathcal{C}_1=-0.999\,998$ is closer to the exact value. On the other hand, Turbiner's and  dVTE trial functions yield accurate values for $\mathcal{C}_0$ and the same corrected value up to six decimal digits.

\begin{table*}[t]
\caption{ Cusp $\mathcal{C}$ and quadrupole moment $Q$ for trial functions Yafet \textit{et al.} (\ref{Yafet}), Turbiner (\ref{Turbiner}), and del Valle \textit{et al.} (\ref{approximatephase}). All numbers were rounded to the digits given.}	
{\setlength{\tabcolsep}{0.175cm}
{\color{black}\begin{tabular}{|c|ccc|ccc|}
		\hline\hline
		\rule{0pt}{3ex}
		Phase & $\mathcal{C}_0$  & $\mathcal{C}_1$ &$\mathcal{C}_0+\mathcal{C}_1$ & $Q_0$& $Q_1$ &  $Q_0+Q_1$  \\[5pt]
		\hline
		\rule{0pt}{4ex}
		Yafet  \textit{et al.}, (\ref{Yafet})
		& \ \ 0.000\,000     &     $-0.999\,998$     &    $-0.999\,998$    &    $0.304\,971$    &    $-0.051\,625$    &    $0.253\,346$       \\[5pt]
		Turbiner, (\ref{Turbiner})
		&    $-1.032\,370$   & \ \  $0.032\,370$     &    $-1.000\,000$    &    $0.558\,045$     &   $-0.168\,602$    &    $0.389\,442$       \\[5pt]
		dVTE, (\ref{approximatephase})
		&    $-0.999\,241$   &    $-0.000\,759$      &    $-1.000\,000$    &    $0.417\,617$     &  \ \ $0.000\,033$  &    $0.417\,650$      \\[5pt]
		\hline\hline
		\end{tabular}}}
		\label{table:expectation}
\end{table*}

Regarding the quadrupole moment, the first-order correction $Q_1$ measures the deviation between the exact and the approximated value provided by Turbiner's and  dVTE trial functions, but not for Yafet's one. Only for the latter function, $Q_1$ actually worsens the estimate.  This indicates that higher-order corrections are needed and that expansion (\ref{exact}) is slowly convergent. Finally, using $Q_0$ and $Q_1$ obtained from (\ref{approximatephase}), we can establish the value of the quadrupole moment, together with its uncertainty, as
\begin{equation*}
Q = 0.417\,62(3),
\end{equation*}
 which captures the exact  value, $Q=0.417\,65$, reported in \cite{Baye2008}.


\section{Conclusions}
\label{Conclusions}	

For a hydrogen atom subjected to a constant magnetic field of strength $\gamma=1$, we estimated the accuracy of the variational energy for three trial functions via the NLP. For all trial functions considered, the first-order correction to the variational energy  $E_2$ delivered an accurate estimate of the deviation  between $E_{var}$ and  the exact energy. This statement was supported by the calculation of $E_3$. We found that solving the relevant equations of the NLP via a collocation method requires only a few collocation points to achieve high accuracy in both $E_2$ and $E_3$. Furthermore, all considered trial functions lead to convergent series whose rates of convergence were estimated. Finally, we showed how the corrections coming from the NLP can be used to estimate the accuracy of any expectation value  calculated by means of the trial functions. As an illustration, we considered the cusp parameter and the quadrupole moment, both defined in terms of expectation values. Additionally, we checked the local accuracy of the trial function proposed in~\cite{TB}.

The numerical implementation based on collocation methods of the NLP can serve as an efficient tool to estimate the accuracy of the variational energy when no reference value is available. Along the same line, and most importantly, this can be used to estimate the accuracy of trial wave functions. Thus, the accuracy of any quantity related to bound states can be investigated (e.g., expectation values, oscillator strengths, relativistic corrections, etc.).

Studying the accuracy of trial functions describing  excited states is also possible in the framework of the NLP. In this case, PT is also developed for the unperturbed nodal surface of the trial function. Then, one estimates and increases the accuracy of the approximate nodal surface using PT corrections. On the one hand, an accurate description of the nodal surface is key to  some methods. For example, it is useful to circumvent the so-called \textit{fermion sign problem} in the context of the fixed-node approximation in Diffusion Monte Carlo.  On the other hand, nodal surfaces can also reveal unexpected symmetries,  as those found for the lowest $^{3}S$ states of He~\cite{Bressanini}. Based on the present results, collocation methods seem promising for studying the nodal surfaces of excited states of few-electron  atomic and molecular systems with high accuracy at low computational cost.


\section*{Acknowledgments}

I am very grateful to A.V. Turbiner  for attracting my attention to the subject, and  also for his insightful discussions and essential remarks on the NLP. I also thank R. Gutierrez-Jauregui and S. Cardenas-Lopez for carefully reading the text and providing valuable suggestions and comments. I appreciate all the support provided by K. Kropielnicka in developing this work. Finally, I thank K. Bartschat for his meticulous reading of the manuscript. Financial support was provided by the Polish National Center for Science (NCN) under Grant No. 2019/34/E/ST1/00390 and the NSF under Grant No. PHY-2110023.


\bibliography{references}
\bibliographystyle{apsrev4-2}


\appendix
\newpage
\section{Matrix Elements}
\label{Aformulas}
The formulae for the matrix elements needed by (\ref{radial}) and (\ref{angular}) are
\begin{align*}
f'_k(r_i) &=
\left\{
\begin{aligned}
\dfrac{r_i}{r_k \left(r_i-r_k\right)}\,\dfrac{L_{Nr}'(r_i)}{L_{Nr}'(r_k)}\ , &\quad i\neq k \cr
\dfrac{r_i+1}{2r_i}\ , &\quad i=k
\end{aligned}
\right.\\
f''_k(r_i) &=
\left\{
\begin{aligned}
\frac{r_i \left(r_i-1\right)- r_k\left(r_i+1\right)}{r_k \left(r_i-r_k\right)^2}\,\dfrac{L_{Nr}'(r_i)}{L_{Nr}'(r_k)}\ , &\quad i\neq k \cr
\dfrac{r_i^2-(N_r-1) r_i-1}{3 r_i^2}\ , &\quad i=k
\end{aligned}
\right.
\end{align*}
and
\begin{align*}
g'_l(u_j) &=
\left\{
\begin{aligned}
\frac{2 u_l}{u_j^2-u_l^2}\,\frac{P_{N_u}'(u_j)}{P_{N_u}'(u_l)}\ , &\quad j\neq l \cr
\frac{1-3 u_j^2}{2 u_j(1- u_j^2)}\ , &\quad j=l
\end{aligned}
\right.\\
g''_l(u_j) &=
\left\{
\begin{aligned}
\frac{4 u_j u_l \left(-3 u_j^2+u_l^2+2\right)}{\left(u_j^2-1\right) \left(u_j^2-u_l^2\right)^2}\ , &\quad j\neq l \cr
\frac{(2 L^2+21) u_j^4-2 \left(L^2+4\right) u_j^2+3}{6 u_j^2 \left(u_j^2-1\right)^2}\ , &\quad j=l
\end{aligned}
\right.
\end{align*}
where $L^2\ =\ N_u(N_u+1)$.


\end{document}